\documentclass[]{aastex631}
\usepackage{graphicx}
\usepackage{color}
\usepackage{epsfig}
\usepackage{subfigure}
\usepackage{amssymb}
\usepackage{amsmath}
\usepackage{natbib}
\usepackage{graphicx}
\usepackage{color}
\usepackage{tabularx}
\usepackage{epsfig}
\usepackage{rotating}
\usepackage{subfigure}
\usepackage{longtable}

\shorttitle{Milky Way Warp Based on Classical Cepheids}
\shortauthors{X. Zhou, X. Chen, L. Deng, S. Wang, J. Wang, J. Zhang}

\graphicspath{ {Figures/} }

\begin{document}

\title{A Detailed Analysis of the Milky Way Warp Based on Classical Cepheids}

\correspondingauthor{Xiaodian Chen}
\email{chenxiaodian@nao.cas.cn}

\author{Xiaoyue Zhou}
\affiliation{CAS Key Laboratory of Optical Astronomy, National Astronomical Observatories, Chinese Academy of Sciences, Beijing 100101, China}
\affiliation{School of Astronomy and Space Science, University of the Chinese Academy of Sciences, Beijing, 100049, China}

\author{Xiaodian Chen}
\affiliation{CAS Key Laboratory of Optical Astronomy, National Astronomical Observatories, Chinese Academy of Sciences, Beijing 100101, China}
\affiliation{School of Astronomy and Space Science, University of the Chinese Academy of Sciences, Beijing, 100049, China}
\affiliation{Institute for Frontiers in Astronomy and Astrophysics, Beijing Normal University, Beijing 102206, China}

\author{Licai Deng}
\affiliation{CAS Key Laboratory of Optical Astronomy, National Astronomical Observatories, Chinese Academy of Sciences, Beijing 100101, China}
\affiliation{School of Astronomy and Space Science, University of the Chinese Academy of Sciences, Beijing, 100049, China}
\affiliation{Department of Astronomy, China West Normal University, Nanchong, 637009, China}

\author{Shu Wang}
\affiliation{CAS Key Laboratory of Optical Astronomy, National Astronomical Observatories, Chinese Academy of Sciences, Beijing 100101, China}
\affiliation{Department of Astronomy, China West Normal University, Nanchong, 637009, China}

\author{Jiyu Wang}
\affiliation{CAS Key Laboratory of Optical Astronomy, National Astronomical Observatories, Chinese Academy of Sciences, Beijing 100101, China}
\affiliation{School of Astronomy and Space Science, University of the Chinese Academy of Sciences, Beijing, 100049, China}

\author{Jianxing Zhang}
\affiliation{CAS Key Laboratory of Optical Astronomy, National Astronomical Observatories, Chinese Academy of Sciences, Beijing 100101, China}
\affiliation{School of Astronomy and Space Science, University of the Chinese Academy of Sciences, Beijing, 100049, China}

\begin{abstract}

Classical Cepheids (CCs) are important probes for the large-scale warp structure of the Milky Way. Using Gaia DR3 CCs, we establish an optimal time-dependent warp model, where the warp height increases with radius following a power-law, the line of nodes (LONs) exhibit linear twisting with radius, following a leading spiral pattern, and the LONs undergo prograde evolution over time. Structurally, we identify significant warp features in the $5-9$ kpc region of the Galactic disk, where the warp model performs better than the flat model. Beyond 15 kpc, the model with the second Fourier term does not fit the observations well, whereas the model with twisted LONs better matches the data. Kinematically, we derived expressions for the vertical velocities using direct differentiation and then calculated the precession rates for each CC. Our results intuitively indicate a nearly uniform and low warp precession rate of  $\omega = 4.86 \pm (0.88)_{stat} \pm (2.14)_{sys}$ km s$^{-1}$ kpc$^{-1}$ beyond 12.5 kpc, in agreement with classical kinematic estimates. Based on these findings, we propose a simple yet comprehensive time-dependent warp model, $Z_{w}(t) = 0.00019R^{3.08}\sin(\phi - (3.87R-41.79 + 4.86t))$, which provides a unified framework for describing both the geometric and kinematic evolution of the Galactic warp. We analyzed the impact of the adopted solar vertical velocity on the inferred warp precession rate and confirmed the reliability of the measured precession rate. In addition, we found that extinction treatment affects the warp amplitude in the inner disk, while its influence on the outer disk warp structure and the precession rate is negligible.

\end{abstract}


\section{Introduction} \label{sec:intro}

Disk warps are a common feature of galaxies, with more than 50\% of spiral galaxy disks warped on the outside \citep{1990MNRAS.246..458S, 1991wdir.conf..181B}. Warp is a large-scale structure of the galaxy disk, which can be simply understood as the disk being made up of a series of tilted rings, each in near-circular motion \citep{2023MNRAS.523.1556D}. For the Milky Way, the disk bends upward along the direction of galactic rotation from nearly the Galactocentric-Sun line, and downward on the other side. The distribution of neutral hydrogen (HI) in the 21-cm radio band first revealed this phenomenon in the Galactic disk \citep{1957AJ.....62...93K,1957Natur.180..677K,1957AJ.....62...90B}, which was later corroborated by various stellar populations \citep{2002A&A...394..883L,2004mim..proc..165Y,2017A&A...602A..67A,2019NatAs...3..320C,2020A&A...637A..96C,2023ApJ...943...88L,2023ApJ...954L...9H,2024MNRAS.527.4863U}, dust \citep{2006A&A...453..635M}, and stellar kinematics \citep{2018MNRAS.481L..21P,2019A&A...627A.150R,2020ApJ...905...49C, 2024A&A...688A..38J}. Classical Cepheids (CCs) have proven to be particularly useful in studying the warp \citep{2023MNRAS.523.1556D,2019NatAs...3..320C,2019Sci...365..478S,2022A&A...668A..40L,2024MNRAS.528.4409C,2024ApJ...965..132Z}. Their well-defined period-luminosity relation allows for precise distance measurements even at large Galactic radii \citep{1912HarCi.173....1L,2012ApJ...758...24F,2022ApJ...938...36R}. Furthermore, as young stars, CCs are minimally affected by complex physical processes such as heating or perturbation in the disk \citep{2008gady.book.....B}, making them ideal tracers for studying the structure and kinematics of the Galactic warp. Despite extensive observations, the origin of the Galactic warp remains uncertain \citep{2025NewAR.10001721H}. Proposed explanations include the misalignment of the dark matter halo with the disk \citep{2009ApJ...703.2068D, 2023NatAs...7.1481H}, persistent misalignment between the disk's rotation axis and normal \citep{1965MNRAS.129..299L}, annexation of smaller galaxies such as the Gaia-Sausage-Enceladus \citep{2024ApJ...975...28D}, accretion of interstellar matter \citep{2002A&A...386..169L}, and interactions with satellite galaxies \citep{2003ApJ...583L..79B, 2006ApJ...641L..33W}. Understanding the warp's formation and evolution is crucial for reconstructing the dynamical history of the Milky Way.

The warp is commonly modeled as a Fourier series in azimuthal angles, where the vertical displacement is described by \citep{2006ApJ...643..881L}:

\begin{linenomath*}
   \begin{equation} \label{eq1}
   \begin{aligned}
     Z_{w}(R,\phi) = Z_{0}+Z_{1}\sin(\phi-\phi_{w1})+Z_{2}\sin(2\phi-\phi_{w2}) \\
   \end{aligned}
   \end{equation}
\end{linenomath*}

Here, $Z_{0}$ denotes the overall height offset of the Galactic disk, the first Fourier term describes the main amplitude feature of the warp, and the second Fourier term represents its lopsidedness. The orientation of the line of nodes (LONs), $\phi_{w1}$, is a crucial parameter in the warp model. Clear evidence of the warp’s north-south asymmetry of the Milky Way has been detected in HI distribution \citep{2006ApJ...643..881L,2008A&A...487..951K}. Based on the HI distribution of external galaxies, it was suggested that the LONs vary with radius and follow a leading spiral pattern \citep{1990ApJ...352...15B}, a trend first identified in the Milky Way using CCs \citep{2019NatAs...3..320C}. However, most studies assume fixed LONs in establishing or utilizing the warp model \citep{2019NatAs...3..320C, 2019AcA....69..305S, 2020ApJ...897..119W, 2022A&A...664A..58C, 2024NatAs...8.1294H}.  
Recent analyses, however, confirm that the LON of the warp is twisted and suggest that $\phi_{w1}(R)$ varies linearly with radius \citep{2023MNRAS.523.1556D, 2024A&A...688A..38J, 2024arXiv240718659P, 2024MNRAS.528.4409C}. Besides, the amplitude of the warp is a function of radius, but its specific form remains uncertain. Generally, it is assumed that the inner region of the Galactic disk is flat, and the warp starts at a certain radius $R_{w1}$. $Z_{1}$ is typically described by either a power-law function, $Z_{1}=a(R-R_{w1})^{b}$ or a linear function $Z_{1}=a(R-R_{w1})$ \citep{2019NatAs...3..320C}. The onset radius of the warp $R_{w1}$ remains uncertain, with studies suggesting that the warp begins near the solar circle \citep{2019Sci...365..478S, 2001ApJ...556..181D, 2018ApJ...864..129H, 2018MNRAS.478.3809S, 2001ASPC..232..229D, 2009A&A...495..819R}. Another function proposed to describe the warp is  $Z_{1}=R\tan(i)$, where $i$ represents the inclination of the inclined plane \citep{2023MNRAS.523.1556D}. Although many studies report significant differences between the northern and southern warp amplitudes \citep{2024MNRAS.527.4863U, 2019A&A...627A.150R}, the second Fourier term is often omitted from models. Observations of CCs indicate a stronger northern warp, leading some studies to incorporate a second Fourier term in the model \citep{2019AcA....69..305S}. At larger radii ($R > 15$ kpc), this term appears to become more significant \citep{2019NatAs...3..320C}. Recently, \cite{2024MNRAS.528.4409C} further confirmed this asymmetry from the maximum amplitudes of the northern and southern warps, and found a constant phase deviation between the LON and the maximum velocity line of the CC warps, reinforcing the need for a second Fourier term. However, discussions of this asymmetry in geometric models remain limited. A comprehensive model that simultaneously incorporates the onset radius, LONs twisting, and warp asymmetry is still needed.

With the release of Gaia Data Release 3 (DR3), a sample of CCs with high-precision distances and line-of-sight velocities (RVs) is now available \citep{2023A&A...674A..17R, 2023A&A...674A..37G}. Based on Galactic CCs, \cite{2023MNRAS.523.1556D} analyzed the Galactic warp by determining the inclination and LON of the disk in each guiding-center radius bin. \cite{2024ApJ...965..132Z} obtained a low precession rate using a classical kinematic approach, and \cite{2024NatAs...8.1294H} pioneered a time-resolved approach to measure the Galactic warp precession rate by tracing variations in the LONs of warps using CCs of different ages. \cite{2024MNRAS.528.4409C} introduced a joint Fourier analysis of the warp heights and vertical velocities of CCs. Unifying the warp's structure and velocity remains challenging. In this study, we aim to identify the optimal model through comparative analysis of different warp models. For the continuum model we use a new method to directly calculate the warp precession rate, which leads to a unified warping evolution model. In Section \ref{sec:data}, we describe the dataset and methodology. In Sect. \ref{sec:result}, we determine the best model based on the comparison. We discuss the observed warp and kinematics in Sect. \ref{sec:4} and \ref{sec:5}. Section \ref{sec:discuss} discusses systematic effects including the solar vertical motion and extinction. Section \ref{sec:concl} summarizes the conclusions.

\section{Data and methods} \label{sec:data}

We use the catalog of Galactic CCs provided by \cite{2023A&A...674A..37G}, which includes a total of 3306 objects. This catalog collected Galactic CCs from the gaiadr3.vari\_cepheid table on the Gaia archive, supplemented with literature samples \citep{2021AcA....71..205P, 2021ApJ...914..127I}. The distances to the CCs were determined using the period-Wesenheit-metallicity (PWZ) relation \citep{2019A&A...625A..14R, 2022A&A...659A.167R}, and the sample was selected to retain only those with a relative distance error of less than 10\% \citep{2023A&A...674A..37G}. We adopted the distance results directly from the catalog. While the distances calculated by Gaia photometry and the mid-infrared photometry exhibit larger biases in the inner regions of the Milky Way, they yield similar results in the outer disk \citep{2025ApJS..278...57S, 2025ApJ...981..179W}. Our study focuses primarily on the warp structure in the outer disk. We assumed that the Sun is in the midplane of the Milky Way ($Z = 0$) and adopted the measured Sun-Galactocentric distance of $R_{\odot}$ = 8277 $\pm$ 9 (stat) $\pm$ 30 (sys) pc \citep{2022A&A...657L..12G}. Using a coordinate transformation method \citep{2023A&A...674A..37G}, we derived ($R, \phi, Z$) for the CCs in the Galactocentric Cylindrical coordinate system. Here, $\phi$ increases in the direction of galactic rotation, with $\phi_{\odot} = 0^\circ$.

Our sample size is small outside of 20 kpc and within 5 kpc, so we first establish a loose criterion for selecting CCs by $5 \leq R \leq 20$ kpc and $|Z| \leq 2$ kpc. In addition, the large extinction toward the Galactic center results in significant obscuration of CCs located on the opposite side of the disk in the midplane, leading to an overrepresentation of CCs situated off the midplane. To mitigate this bias, we exclude CCs with $120^\circ < \phi < 240^\circ$, leaving a sample of 3037.

We performed an initial model fit to the CC samples to exclude outliers based on residual analysis. The model adopted is the classical warp model, where the inner disk is flat and the warp height of the outer disk follows a power-law dependence on radius. Specifically, for $R > 9$ kpc, the warp height is given by $Z_{w} = a(R-9)^{b}\sin(\phi - \phi_{w})+Z_{0}$, while for $R \leq 9$ kpc, $Z_{w}=Z_{0}$. For each CC, we take its vertical Galactocentric coordinate $Z_{i}$ as the observational warp height $Z_{w,i}^{\mathrm{obs}}(R_{i},\phi_{i})$. We use the \texttt{scipy.optimize.curve\_fit} routine to perform a least squares fit and obtain the optimal warp parameters. By minimizing the residual sum of squares, $\sum_{i=1}^{N}\left[Z_{w,i}^{\mathrm{obs}}(R_i,\phi_i)-Z_{w,i}^{\mathrm{model}}(R_i,\phi_i)\right]^2$, we thereby derive the functional form of $Z_{w}^{\mathrm{model}}$. The analytic warp model introduced in Sect. \ref{sec:result} is fitted with the same approach. We directly analyzed the distributions of the residuals $\Delta Z_{w,i}=Z_{w,i}^{\mathrm{obs}}-Z_{w,i}^{\mathrm{model}}$. At each radius, some samples clearly deviate from the model. We applied a radius-dependent filtering method, where the threshold was guided by the Galactic flare height \citep{2019NatAs...3..320C}, which increases with radius. The right panels of Fig. \ref{A1.fig}-\ref{A3.fig} present the residual thresholds, encompassing the vast majority of stars. Through threshold filtering, we excluded 211 outliers, while preserving the key structural features of the sample. Notably, we detect a clear asymmetry in the residuals within the 5–9 kpc range (see Fig. \ref{A1.fig} in Appendix \ref{Appendix}), consistent with the expected Galactic warp signature. This implies the presence of warps in these radius ranges, which we explore in detail later. Additionally, we examined the residual distribution as a function of radius across different azimuthal slices, identifying and removing one further clear outlier. Our final geometric sample includes 2826 CCs. 

In Sect. \ref{sec:5}, we utilize the Gaia DR3 kinematic data of CCs, which we describe here. We obtained multi-dimensional astrometric data for the CCs by cross-matching the gaiadr3.vari\_cepheid catalog with the gaiadr3.gaia\_source catalog from the Gaia archive. Then we cross-matched this resulting catalog with the Galactic CC catalog to retain only stars within the Milky Way. Following the recommendation of \cite{2023A&A...674A..37G}, we retained only stars with mean RV estimates derived from the spectroscopic pipeline, resulting in a total of 2057 stars. For this sample, we extracted right ascension, declination, proper motions, RVs, and distances derived from the PWZ relation. These parameters were used to compute the three-dimensional (3D) velocities in the Galactic cylindrical polar coordinate system (for details on the coordinate transformation methodology, refer to Sect. 3.2 of \cite{2023A&A...674A..37G} and Sect. 2.2 of \cite{2024ApJ...965..132Z}). The solar motion adopted is $[U_{\odot}, V_{\odot}, W_{\odot}] = [9.3 \pm 1.3,\ 251.5 \pm 1.0,\ 8.59 \pm 0.28]\ $ km s$^{-1}$ \citep{2023A&A...674A..37G}. We applied velocity constraints of $V_{Z}$ in [$-50, 50$] km s$^{-1}$, $V_{R}$ in [$-100, 100$] km s$^{-1}$ and $V_{\phi}$ in [$160, 300$] km s$^{-1}$ to exclude outliers. Additionally, we binned the Galactocentric radius with a bin width of 1 kpc and removed outliers using a 3$\sigma$ clipping procedure. Our final kinematic sample consists of 1939 CCs.

\section{Warp Model} \label{sec:result}

We used six different models to investigate the structure of the Galactic warp.

\begin{enumerate}
    \item \emph{Power-law model} --- This model describes a continuous warp structure, where the warp height increases smoothly as a power-law function of radius: $Z_{w} = aR^{b}\sin(\phi-\phi_{w})$. 
    \item \emph{Power-law and $R_{1}=9$ \rm{kpc} model} --- This follows the classical warp model, assuming a flat inner disk with the warp amplitude growing beyond a fixed onset radius: \\ $Z_{w} =
    \begin{cases}
    a(R-9)^{b}\sin(\phi-\phi_{w})+Z_{0}, & R > 9 \text{ kpc} \\
    Z_{0}, & R \leq 9 \text{ kpc}
    \end{cases}$.   
    \item \emph{Linear and $R_{1}=9$ \rm{kpc} model} --- Similar to the previous model, but assuming a linear increase in warp amplitude beyond the onset radius:
    $Z_{w} =
    \begin{cases}
    a(R-9)\sin(\phi-\phi_{w}) + Z_{0}, & R > 9 \text{ kpc} \\
    Z_{0}, & R \leq 9 \text{ kpc}
    \end{cases}$. 
    \item \emph{Linear and $R_{1}=9,\ R_{2}=15$ \rm{kpc} model} --- This model introduces asymmetry beyond $R = 15$ kpc by incorporating a second Fourier term to describe the differential behavior of the warp at large radii: \\
    $Z_{w} =
    \begin{cases}
    Z_{0}, & R \leq 9 \text{ kpc} \\
    Z_{0} + a(R-9)\sin(\phi-\phi_{w}) , & 9 < R \leq 15 \text{ kpc} \\
    a_2(R-15) \sin(2\phi-\phi_{w2}) + a(R-9)\sin(\phi-\phi_{w}) + Z_{0} , & R > 15 \text{ kpc}
    \end{cases}$. 
    \item \emph{Power-law with twisted model} --- This extends the power-law model by allowing the LONs to twist with radius, incorporating both an exponential warp structure and a precession term: 
    $Z_{w} = aR^{b}\sin(\phi - \phi_{w}(R)), \quad \phi_{w}(R) = cR + d$.  
    \item \emph{Linear and $R_{1}=9$ \rm{kpc} with twisted model} --- A linear version of the previous model, assuming a flat inner disk and a twisting LONs beyond the onset radius: \\
    $Z_{w} =
    \begin{cases}
    Z_{0}, & \text{if } R \leq 9 \\
    a(R-9)\sin(\phi - \phi_{w}(R)) + Z_{0}, & \text{if } R > 9
    \end{cases}$. 
\end{enumerate}

We adopt 9 kpc as the onset radius for the CC warp, a widely accepted value in the literature. Since the warp amplitude parameters a, b and $R_{w}$ are inherently coupled, varying the onset radius affects a and b. However, these parameters collectively define the warp amplitude, and the overall warp structure remains largely independent of the chosen onset radius. To verify this, we tested an alternative onset radius of 7 kpc, which showed no impact on regions within \( R < 7 \) kpc or beyond \( R > 10 \) kpc, with only a slight reduction in the root mean square error (RMSE) for the \( 7-9 \) kpc range.  In this study, we focus on segmented vs. continuous function models, models with single vs. double Fourier terms, and non-twisted vs. twisted warp models. Given these considerations, adopting \( R_{1} = 9 \) kpc as the onset radius remains an appropriate choice.

We found only 11 data points beyond 19 kpc, making the fit in this region unreliable. Therefore, we used 2815 CCs with $R < 18.5$ kpc for the model fitting. The least-squares fitting method was applied to all six types of models to obtain the best-fit parameters, and the results are shown in Table \ref{tab:my_label}.

\begin{table}
 \caption{Fitting parameters of the warp model. The function expressions for each model are provided in Section \ref{sec:result}. The best-fit parameters and their $1\sigma$ standard deviations, obtained through least-squares fitting, are listed below. The last column lists the overall RMSE values for each model fit.}
    \centering
    \renewcommand{\arraystretch}{1.5}
        
    \resizebox{\textwidth}{!}{
    \begin{tabular}{
    lccccccc}
    \toprule 
     Model  & $Z_{0}$ (kpc) & a & b & $\phi_{w}\ (^\circ)$ & $a_{2}$ & $\phi_{w2}\ (^\circ)$ & RMSE (kpc) \\
      \hline
    \emph{Power-law model}
     &  & 0.000153$\pm$0.000029 & 3.17$\pm$0.07  & 13.88$\pm$0.65 &  &  & 0.197\\
    \emph{Power-law and $R_{1}=9$ model}  & -0.0134$\pm$0.0842 & 0.133$\pm$0.008 & 1.03$\pm$0.04 & 12.78$\pm$0.80 &  & &0.199\\
    \emph{Linear and $R_{1}=9$ model}
     & -0.0134$\pm$0.0842 & 0.140$\pm$0.002 & 1 & 12.70$\pm$0.80 &  & &0.199\\
    
    \emph{Linear and $R_{1}=9,\ R_{2}=15$ model}
     & -0.0134$\pm$0.0842 & 0.141$\pm$0.002 & 1 & 4.23$\pm$0.93 & -0.083$\pm$0.020 & -89.28$\pm$13.44 &0.203\\
    
    \emph{Power-law with twisted model}
    & & 0.00019$\pm$0.00003 & 3.08$\pm$0.07 & (3.87$\pm$0.27)R-(41.79$\pm$3.95) &  &  &0.190\\
   
    \emph{Linear and $R_{1}=9$ with twisted model}
     & -0.0134$\pm$0.0842 & 0.140$\pm$0.002 & 1 & (5.57$\pm$0.41)R-(68.61$\pm$6.03) &  & &0.190\\
         \hline
    \end{tabular}
    }

    \label{tab:my_label}

\end{table}

\begin{figure*}
\centering
  \begin{minipage}{\textwidth}
  \includegraphics[width=\linewidth]{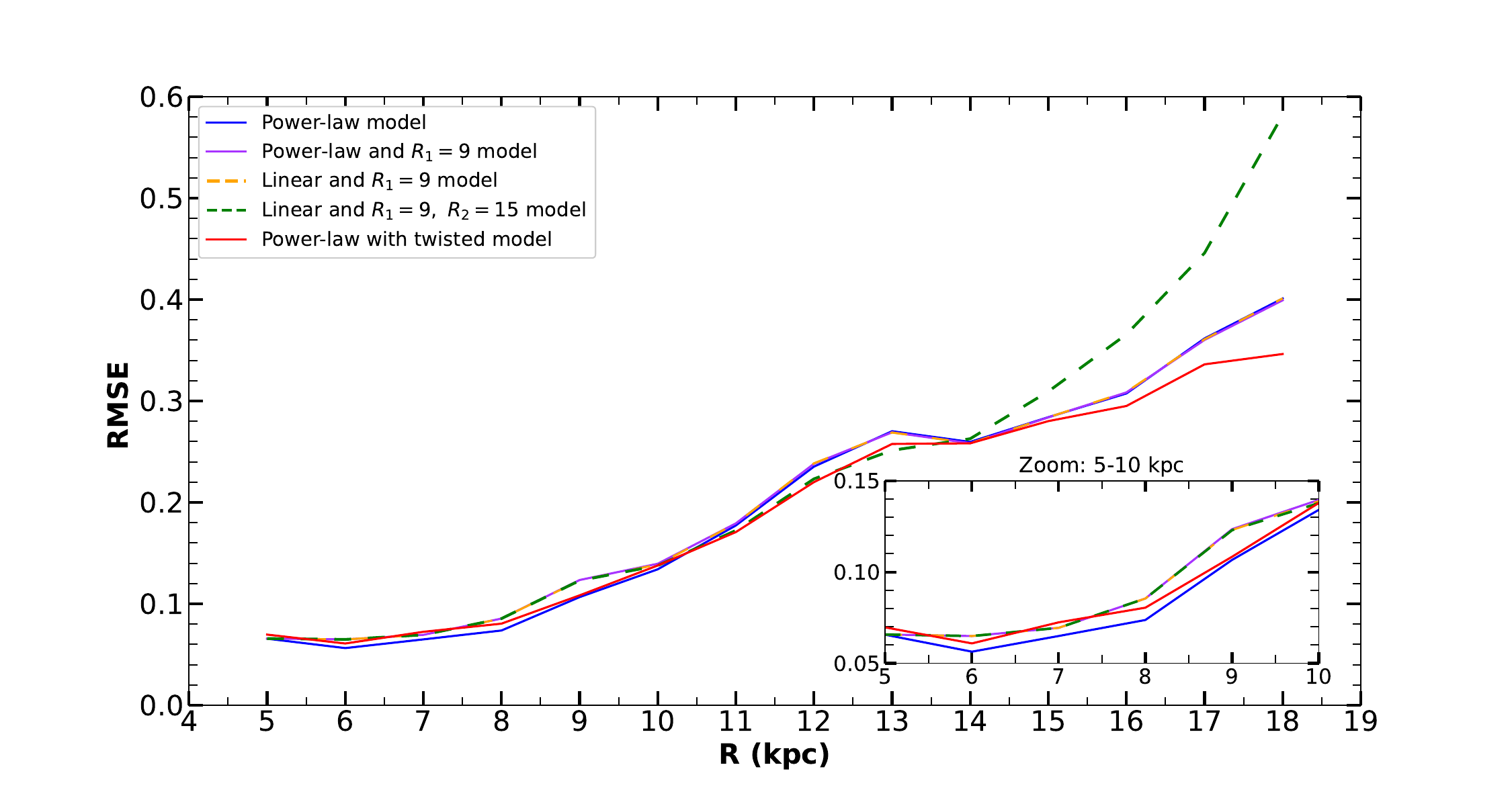}
\caption{Comparison of RMSE for different models. The figure presents five models, described in detail in the main text, represented by the blue solid line, bright purple solid line, orange dashed line, green dashed line, and red solid line. The variation of the RMSE with radius is shown for each model. The bottom-right subplot provides a zoomed-in view of the results for the $5-10$ kpc range.}\label{1.fig}
  \end{minipage}
  \end{figure*}

To determine the best-fit model, we divided the radius into 1 kpc intervals and calculated the RMSE values for each model. Fig. \ref{1.fig} shows the variation of RMSE with radius for different models, represented by different colors and symbols. The RMSE increases with radius, reflecting greater dispersion in the outer disk, consistent with Galactic flare. Within $R<10$ kpc, the \emph{Power-law model} (blue solid line) provides the best fit. The \emph{Power-law and $R_{1}=9$ model}, \emph{Linear and $R_{1}=9$ model}, and \emph{Linear and $R_{1}=9,\ R_{2}=15$ model}, all assuming a flat disk within 9 kpc, have the same RMSE, which is slightly higher than that of the \emph{power-law model}. This suggests that models with the warp at smaller radii ($R_{1}<9$ kpc) are better suited than flat models. Beyond 10 kpc, the \emph{Power-law with twisted model} (red solid line) exhibits the lowest RMSE, significantly outperforming others. The \emph{Linear and $R_{1}=9,\ R_{2}=15$ model} (green dashed line), which includes warp asymmetry beyond 15 kpc, shows a larger RMSE beyond 14 kpc, indicating that the second Fourier term does not improve the fit. Fig. \ref{B1.fig} compares the RMSEs of two twisted models (\emph{Power-law with twisted model} and \emph{Linear and $R_{1}=9$ with twisted model}) with the non-twisted \emph{Power-law model}. The results indicate that twisted models outperform the non-twisted model, and the warped model performs better than the flat model for $R<9$ kpc. Thus, our optimal model is the \emph{Power-law with twisted model}, given by $Z_{w} = (0.00019\pm0.00003)R^{3.08\pm0.07}\sin(\phi - ((3.87\pm0.27)R - (41.79\pm3.95)))$. 

\begin{figure*}
\centering
  \begin{minipage}{\textwidth}
  \includegraphics[width=\linewidth]{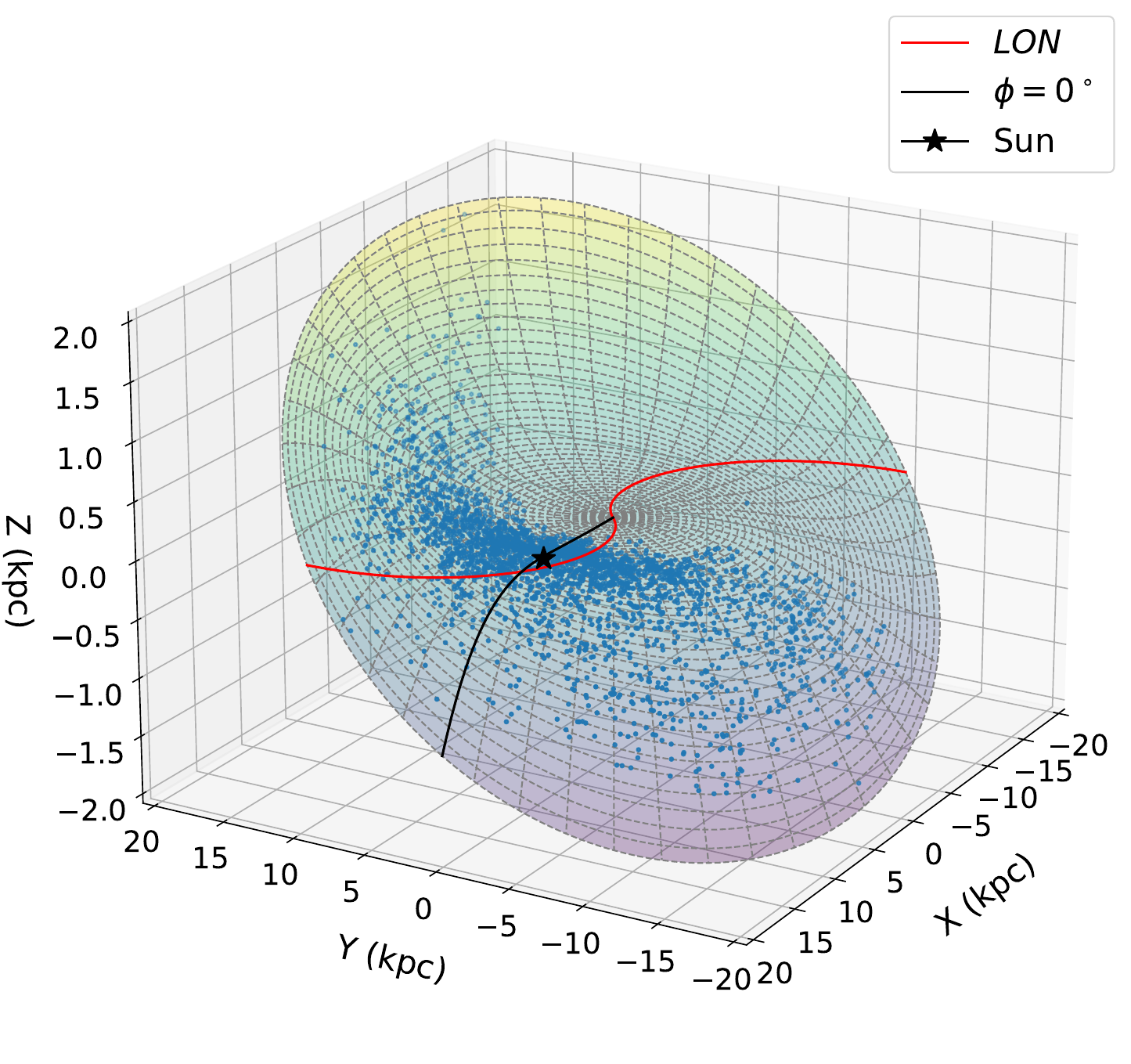}
\caption{Visualization of the best-fit 3D model, the \emph{Power-law with twisted model}. The grid lines represent the 2D surface of the mid-plane, illustrating the warp structure. Lighter and darker regions indicate higher and lower warp heights, respectively. The line of nodes (LONs) and the Galactocentric-Sun line are shown as solid red and black lines, respectively. The Sun’s position is marked with a pentagram, while the Galactic Center represents the disk's center. The CC observations are shown as blue dots.} \label{2.fig}
  \end{minipage}
  \end{figure*}

Fig. \ref{2.fig} presents the 3D warp structure, where the LONs clearly deviate from the Galactocentric-Sun direction, forming a leading spiral pattern, consistent with Briggs' rule for spiral galaxies \citep{2023MNRAS.523.1556D, 2019NatAs...3..320C, 2024MNRAS.528.4409C, 1990ApJ...352...15B, 2024arXiv240718659P}. Comparing our results with the latest CC geometric warp model \citep{2024arXiv240718659P}, their model gives a warp onset radius of $7.7\pm0.4(stat)\pm0.1(sys)$ kpc, with flat LONs of $\phi_{LON,0} = 0.9\pm1.1(stat)\pm0.4(sys)$ within 12.6 kpc, increasing beyond this radius at a rate of $\beta_{w} = 8.0\pm0.5(stat)\pm0.2(sys)$ deg/kpc. Adopting their model and fitting our sample, we obtained $\phi_{LON,0} = -0.9\pm1.1(stat)$ deg within 12.6 kpc and a rate of $\beta_{w} = 6.7$ deg/kpc, consistent within 1$\sigma$ errors. The determined RMSE is nearly identical to that of our model. These findings confirm the reliability of our sample and results. Compared to their model, our model is simpler as it does not assume the onset radius of the warp and the radius at which the LONs start to change.

\section{Warp in Radial and Azimuthal Slices} \label{sec:4}
To thoroughly evaluate the performance of different warp models, we analyze their fit to the observational data in both radial and azimuthal slices, examining variations in warp height as a function of radius and Galactic azimuth angle.

\begin{figure*}
\centering
  \begin{minipage}{\textwidth}
  \includegraphics[width=\linewidth]{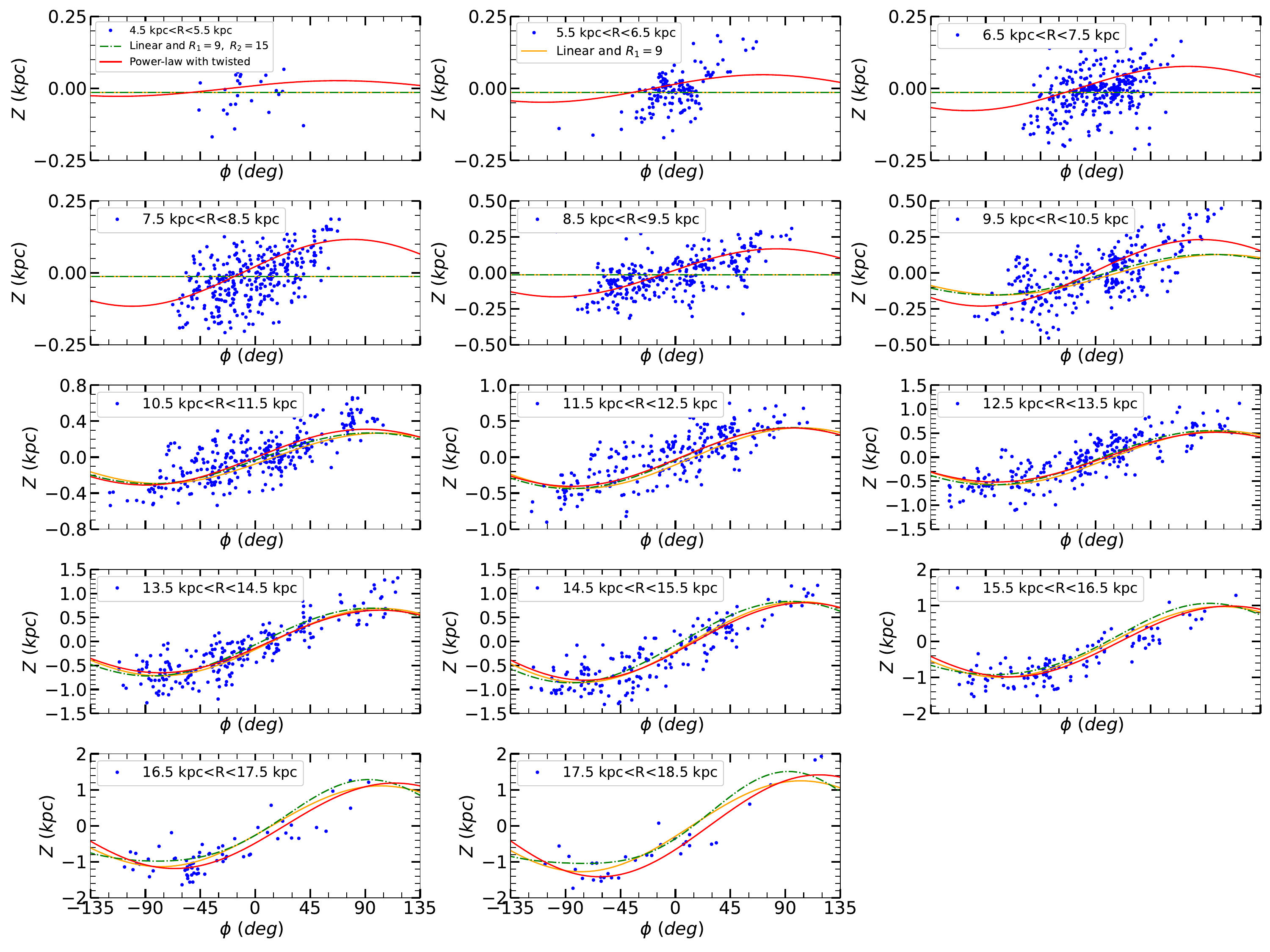}
\caption{Fitting performance of the models in radial slices within the range of $5-18$ kpc. The x-axis represents azimuth and the y-axis represents height. Blue dots show the distribution of the sample data in the $\phi-Z$ plane at each radius. The orange line, green dot-dashed line, and red line correspond to the \emph{Linear and $R_{1}=9$ model}, \emph{Linear and $R_{1}=9,\ R_{2}=15$ model} and \emph{Power-law with twisted model}, respectively. \label{3.fig}}
  \end{minipage}
  \end{figure*}

In Fig. \ref{3.fig}, we compare the fitting performance in radial slices using three representative models: \emph{Linear and $R_{1}=9$ model}, \emph{Linear and $R_{1}=9,\ R_{2}=15$ model} and \emph{Power-law with twisted model}. They represent the commonly used segmental warp model, the warp model with asymmetric outer disk, and our optimal twisted model, respectively. In Fig. \ref{3.fig}, the \emph{Power-law with twisted model} (red solid line) provides the best fit, revealing an asymmetric warp height distribution for $R>5.5$ kpc. All three models agree well in the $11-14$ kpc range, capturing the main warp features. Beyond 15 kpc, the \emph{Linear and $R_{1}=9,\ R_{2}=15$ model} (green dot-dashed line) shows a significant asymmetry between the northern and southern warp, which is inconsistent with our sample, where height asymmetry is weak. Additionally, the \emph{Linear and $R_{1}=9$ model} (orange solid line) begins to diverge from the \emph{Power-law with twisted model} particularly in azimuthal variations at peak warp heights.

\begin{figure*}
\centering
  \begin{minipage}{\textwidth}
  \includegraphics[width=\linewidth]{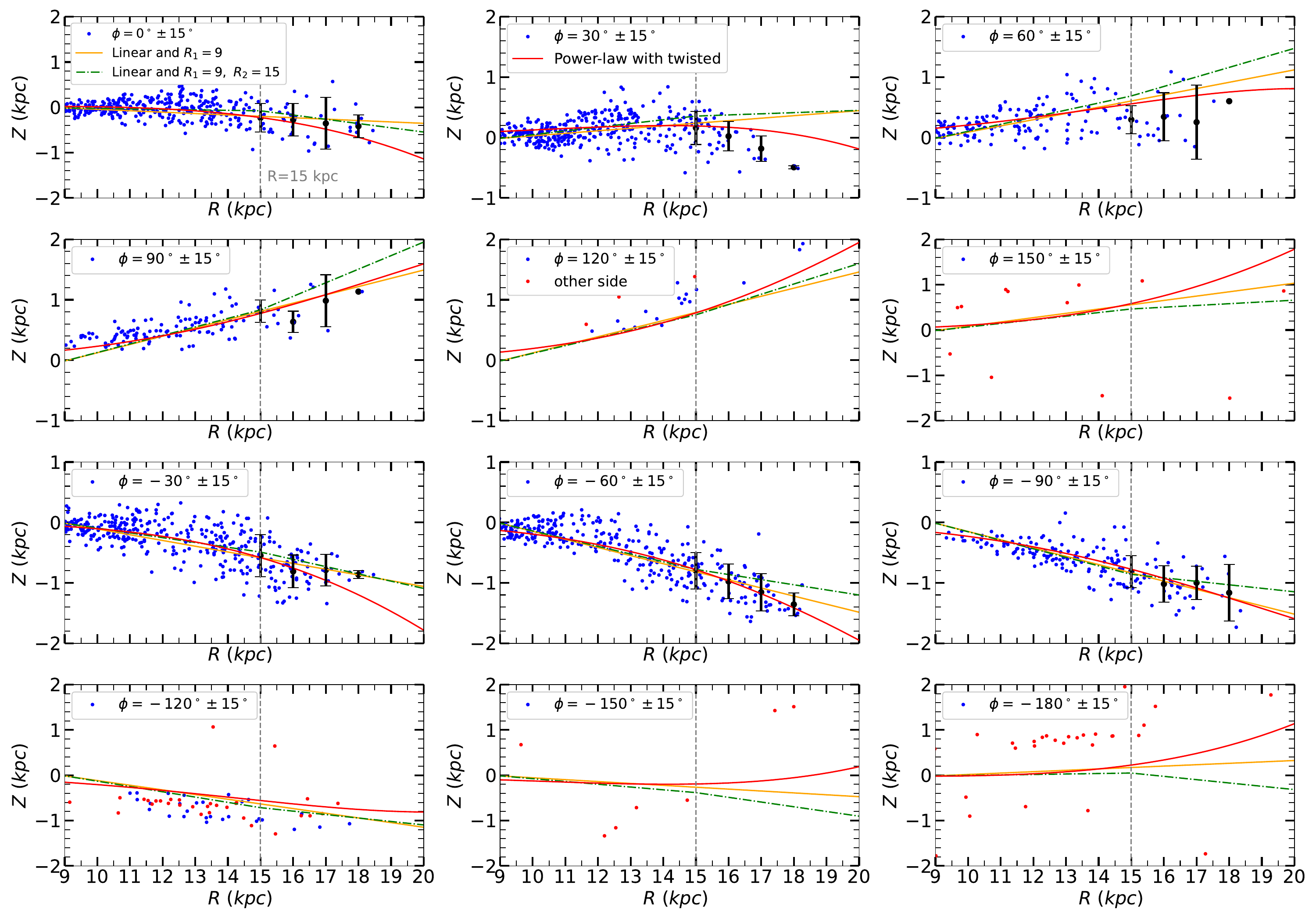}
\caption{Fitting results of the three models in different azimuthal slices. Each panel shows the variation of CC height with radius within a specific azimuthal region. The \emph{Power-law and $R_{1}=9$ model}, the \emph{Linear and $R_{1}=9,\ R_{2}=15$ model}, and the \emph{Power-law with twisted model} are represented by orange line, green dashed line, and red line, respectively. Blue dots represent the CC sample, while red dots correspond to samples in the $120^\circ$ to $180^\circ$ and $-180^\circ$ to $-120^\circ$ azimuthal regions. The mean height and $1\sigma$ dispersion beyond 15 kpc are marked with black dots.} \label{4.fig}
  \end{minipage}
  \end{figure*}

To further investigate the azimuthal dependence of the warp, we divide the disk into 12 regions in $30^\circ$ bins, spanning from $-180^\circ$ to $180^\circ$. Fig. \ref{4.fig} shows CC heights versus radius for each azimuthal region, with the three models overlaid. We also include samples from the far side of the Galactic center ($120^\circ$ -- $180^\circ$ and $-120^\circ$ -- $-180^\circ$) to examine the structure. These samples are mainly from regions above or below the midplane due to extinction effects (red dots). For a clearer comparison of the differences between the models and observations beyond 15 kpc, we plot the mean height and $1\sigma$ dispersion at each radius in the azimuthal zones of $0^\circ$, $\pm30^\circ$, $\pm60^\circ$, and $\pm90^\circ$, shown in black color in Fig. \ref{4.fig}. Beyond 15 kpc (gray dashed lines), the northern Galactic disk ($30^\circ$ -- $60^\circ$) shows a lower mean height than the \emph{Linear and $R_{1}=9$ model} (orange solid line), with discrepancies decreasing around $90^\circ$. Conversely, the southern Galactic disk ($-30^\circ$ to $-60^\circ$) exhibits a slightly higher mean height than this model. The \emph{Linear and $R_{1}=9$ model} assumes a linear height variation with radius in each azimuthal slice, but CCs do not follow this trend, particularly at $30^\circ$ and $-60^\circ$, indicating that classical linear warp models fail to capture observed features. Beyond 15 kpc, the \emph{Linear and $R_{1}=9,\ R_{2}=15$ model} (green dot-dashed line) exhibits excessive asymmetry, with an amplified northern warp and a suppressed southern warp, deviating from the data. In contrast, the \emph{Power-law with twisted model} (red solid line) aligns closely with observations. This suggests that between 15 and 18 kpc, the warp's LONs undergo a prograde twist, transferring warp features from the southern to the northern Galactic disk. Observational evidence includes the downward-bending warp outside the Galactic anticenter direction  ($0^\circ-30^\circ$) and the enhanced warp amplitude in the southern Galactic disk ($-30^\circ - -60^\circ$).

\section{Warp Precession Rate}
\label{sec:5}

The warp precession rate $\omega$ describes the evolution of the warp LONs over time, offering insights into the disk's vertical kinematics. Classical warp kinematic model derives $\omega$ from the zeroth moment of the collisionless Boltzmann equation \citep{2020NatAs...4..590P}. Here, we adopt a new method to derive the expression of the kinematic model based on our best-fit geometric warp model—the \emph{Power-law with twisted model}—in which CC positions evolve with time as ($R(t), \phi(t), Z(t)$). The LON varies as: $\phi_{w} = \phi_{0,w} + \omega t$, $\phi_{0,w}(R) = cR + d$. The current LONs are twisted with radius but time-independent. Thus, it is assumed that the motion of the stars in the disc follows the shape of the warp:
\begin{linenomath*}
   \begin{equation} \label{eq2}
   \begin{aligned}
     Z_{w}(t) = aR(t)^{b}\sin(\phi(t) - (\phi_{0,w}(R)+\omega t)), \quad \phi_{0,w}(R) = cR + d. \\
   \end{aligned}
   \end{equation}
\end{linenomath*}
Differentiating this expression, and using $\frac{dR(t)}{dt} = V_{R}(t)$, $\frac{d\phi(t)}{dt} = \frac{V_{\phi}(t)}{R}$, we obtain:  
\begin{linenomath*}
   \begin{equation} \label{eq3}
   \begin{aligned}
   V_{Z}(t) = abR(t)^{b-1}V_{R}(t)\sin(\phi(t) - (\phi_{0,w}+\omega t))+\left( \frac{V_{\phi}(t)}{R}-\omega \right)\ aR(t)^{b}\cos(\phi(t) - (\phi_{0,w}+\omega t)) \\
   \end{aligned}
   \end{equation}
\end{linenomath*}
For the present observations $t = 0$, this simplifies to: 
\begin{linenomath*}
   \begin{equation} \label{eq4}
   \begin{aligned}
   V_{Z}(t=0) = abR^{b-1}V_{R}\sin(\phi - \phi_{0,w})+\left( \frac{V_{\phi}}{R}-\omega \right)\ aR^{b}\cos(\phi - \phi_{0,w}) \\
   \end{aligned}
   \end{equation}
\end{linenomath*}
This expression has the same structure as classical warp kinematic models \citep{2020ApJ...905...49C, 2020NatAs...4..590P}, but is derived directly from a differentiable geometric function, thus providing a more general formulation. For comparison, the model in \cite{2020NatAs...4..590P}, which allows for time-varying warp amplitude, is:
\begin{linenomath*}
\begin{equation} \label{eq1:Poggio2020}
\begin{aligned}
\overline{V}_{z}(R,\phi,t=0) = 
\left( \frac{\overline{V}_{\phi}}{R} - \omega(R) \right) 
      h_{w}(R)\,\cos(\phi - \phi_{w})  + \frac{\partial h_{w}}{\partial t}\sin(\phi - \phi_{w}).
\end{aligned}
\end{equation}
\end{linenomath*}
Substituting our \emph{Power‐law with twisted model}, with $h_{w}(R) = aR(t)^{b}$  and $\frac{\partial h_{w}}{\partial t}=abR(t)^{b-1}\overline{V}_{R}$, into this equation gives the following kinematic model: 
\begin{linenomath*}
\begin{equation} \label{eq2:Poggio2020}
\begin{aligned}
\overline{V}_{z}(R,\phi,t=0) = 
\left( \frac{\overline{V}_{\phi}}{R} - \omega(R) \right) 
      aR^{b}\,\cos(\phi - \phi_{w})  + abR^{b-1}\overline{V}_{R}\sin(\phi - \phi_{w}).
\end{aligned}
\end{equation}
\end{linenomath*}
Similarly, the formulation by \cite{2020ApJ...905...49C}, which incorporates radial velocity effects, is:
\begin{linenomath*}
\begin{equation} \label{eq1:Cheng2020}
\begin{aligned}
\overline{V}_{z} &= 
\left( \frac{\overline{V}_{\phi}}{R} - \omega \right)\,
      h(R)\cos\theta
      + \overline{V}_{R}\,\frac{dh}{dR}\sin\theta, \theta=\phi - \phi_{w}
\end{aligned}
\end{equation}
\end{linenomath*}
For our \emph{Power‐law with twisted model} with $h(R) = aR^{b}$  and $\frac{dh}{dR}=abR^{b-1}$, the kinematic model is:
\begin{linenomath*}
\begin{equation} \label{eq2:Cheng2020}
\begin{aligned}
\overline{V}_{z} &= 
\left( \frac{\overline{V}_{\phi}}{R} - \omega \right)\,
      aR^{b}\cos\theta
      + \overline{V}_{R}\,abR^{b-1}\sin\theta,\ \theta=\phi - \phi_{w}
\end{aligned}
\end{equation}
\end{linenomath*}
These comparisons show that the classical kinematic warp prescriptions are special cases of our general differentiable formulation, where the vertical velocity field is obtained via direct time differentiation of the warp surface.

To derive $\omega$, we first supplemented velocity data by adopting the solar motion velocity $[U_{\odot},V_{\odot},W_{\odot}] = [9.3\pm1.3, 251.5\pm1.0, 8.59\pm0.28]$ km s$^{-1}$ \citep{2023A&A...674A..37G}, calculated the 3D velocities ($V_{R}$, $V_{\phi}$, $V_{Z}$), and excluded outliers exceeding 3 $\sigma$ based on velocity distributions (see Sect. \ref{sec:data}). Cross-matching geometric and kinematic samples, we obtained 1855 CCs with reliable velocity data. Using Monte Carlo (MC) simulations, we generated 2000 random values per CC, accounting for uncertainties in distance, proper motion, and RV.

\begin{figure*}
\centering
  \begin{minipage}{\textwidth}
  \includegraphics[width=\linewidth]{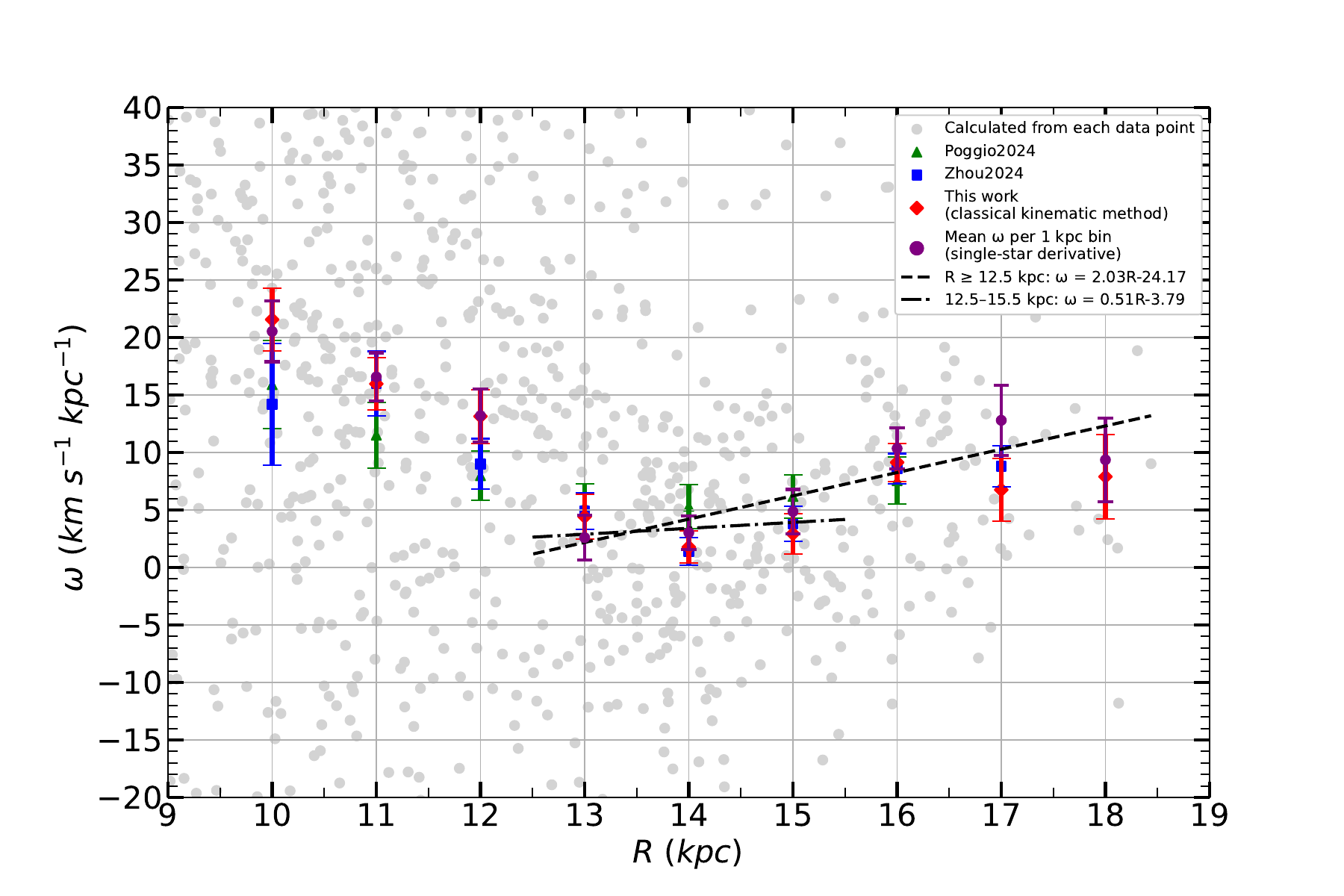}
\caption{Comparison of the warp precession rate variation with radius. Red diamonds show the results obtained by applying the \emph{Power-law with twisted model} to the classical kinematic method, with error bars indicating the $1\sigma$ standard deviation. Blue squares and green triangles show results from \cite{2024ApJ...965..132Z} and \cite{2024arXiv240718659P}, respectively. Light gray points indicate precession rates computed for individual CCs using direct differentiation. Purple circles show the binned mean $\omega$ values (1 kpc radial bins), with error bars representing SEM. Black lines show linear fits to gray points at $R \geq 12.5$ kpc (dashed) and $R=12.5-15.5$ kpc (dash-dotted). } \label{5.fig}
  \end{minipage}
  \end{figure*}

We calculated the warp precession rate using two approaches: (1) Least-squares fitting, which estimates the mean precession rate at each radius based on velocity distributions, providing a statistical measure (using Eqs. \ref{eq2:Poggio2020} and \ref{eq2:Cheng2020}); and (2) Direct calculations for individual CCs using Eq. \ref{eq4}, allowing us to explore radial variations in $\omega$. For the least-squares method, we combined the the optimal geometric warp model, the classical kinematic warp model and observed velocity data. We performed 10,000 MC iterations on the variables $R$, $\phi$, $V_{R}$, $V_{\phi}$, and $V_{Z}$ to estimate uncertainties. The final precession rate was obtained as the average across all simulations at each radius, $1\sigma$ uncertainties reflecting both fitting errors and the dispersion of $\omega$ (red diamonds in Fig. \ref{5.fig}). Our results agree within $1\sigma$ with previous CC-based warp studies \citep{2024ApJ...965..132Z,2024arXiv240718659P}, confirming the suitability of our best-fit geometric model for kinematic analyses.

In parallel, we computed the precession rate of each CC directly using Eq. \ref{eq4}. This revealed the distribution of $\omega$ across radius (light gray dots in Fig. \ref{5.fig}). To ensure robustness, we applied MC simulations and retained only stars with precession rate errors below 10 km s$^{-1}$ kpc$^{-1}$. The results show large dispersion within 12 kpc but significant stabilization beyond 12.5 kpc, where the precession rate converges to $4.86 \pm (0.88)_{stat} \pm (2.14)_{sys}$ km s$^{-1}$ kpc$^{-1}$. We conducted a nested MC to obtain robust results and statistical errors. The inner layer randomly sampled the parameter 2000 times for each star and recorded the mean precession rate of the subsample with $R > 12.5$ kpc along with the standard error of that mean (SEM). The outer layer repeated the entire inner procedure 100 times, producing a set of precession rates. We took the average of this set as the precession rate, and its statistical uncertainty is obtained by quadratically adding the average SEM to the scatter of the 100 precession rates. The systematic uncertainties are addressed in Sect. \ref{sec:6-3}. 

The means and statistical uncertainties per radial bin derived from the direct derivative method are shown as purple points in Fig. \ref{5.fig}, closely matching the least-squares results and confirming consistency between the two methods. Although $\omega$ appears slightly elevated within 12 kpc, it remains within uncertainties, suggesting only minor influence from local disk structures. While the least-squares method offers a statistical perspective based on the collisionless Boltzmann equation, the direct approach provides a more intuitive view by linking individual stellar motions to global warp dynamics. As shown in Fig. \ref{5.fig}, we directly observe how CC kinematics contribute to the evolution of the Galactic warp, without relying solely on bulk velocity distributions.

To explore the radial dependence of $\omega$, we performed two linear fits in Fig. \ref{5.fig}. The dashed black line, fitted over all points with $R > 12.5$ kpc, yields $\omega = 2.03R - 24.17$ km s$^{-1}$ kpc$^{-1}$, indicating a monotonic increase in $\omega$ with radius and supporting a leading spiral pattern of the LONs. A restricted fit within $12.5 \leq R \leq 15.5$ kpc (dot-dashed line) gives $\omega = 0.51R - 3.79$ km s$^{-1}$ kpc$^{-1}$, consistent with solid-body precession. The increasing trend beyond 15.5 kpc suggests additional torques—potentially from a misaligned halo or recent accretion—may accelerate warp precession in the outer disk.

By combining individual (gray) and ensemble (red) precession rates, we adopt the statistical mean of all stars at $R > 12.5$ kpc as the global warp precession rate: $\omega = 4.86 \pm (0.88)_{stat} \pm (2.14)_{sys}$ km s$^{-1}$ kpc$^{-1}$. Although the fitted $\omega$ is slightly higher within 12 kpc, the deviation remains within error bars. Finally, we propose a time-dependent warp model based on our kinematic analysis: $Z_{w}(t) = 0.00019R^{3.08}\sin(\phi - (3.87R-41.79 + 4.86t))$. This model offers a kinematic framework for visualizing Galactic warp evolution. When integrated with simulations, it can aid in reconstructing the long-term dynamical history of the Milky Way. Future work will apply this model to tracers of various ages to probe different epochs of warp evolution.

\section{Discussion}
\label{sec:discuss}
In this section, we focus on discussing current controversies in the structural and kinematic studies of the Galactic warp. Recent analyses have highlighted two key sources of uncertainty: the adopted reference value for the solar vertical velocity $V_{Z,\odot}$ and the impact of extinction on distance measurements. We specifically examine how these two factors influence our determination of the geometry and precession rate of the Galactic warp. Finally, we analyzed the systematic uncertainties in the warp precession rate arising from the various factors.

\subsection{Impact of Solar Vertical Velocity on Warp Precession Rate}
\label{sec:6-1}

As shown in Fig. \ref{3.fig}, the warp is already evident at the solar radius of 8 kpc, suggesting that the adopted reference value of the Sun’s vertical motion, $V_{Z,\odot}$, may be underestimated. \cite{2024arXiv241220344P} pointed out that the $0.6^\circ$ inclination of the local solar orbital plane relative to the Galactic plane projects the Sun’s rotational velocity onto the vertical direction, resulting in a vertical velocity component of $V_{Z,\odot} \approx 2$ km s$^{-1}$. Neglecting this vertical component leads to a systematic underestimation of the warp tilt angle derived from orbital angular momentum, compared to that obtained from 3D structure. It also produces a mismatch in the measured LONs between the two approaches. To investigate this, we quantify the impact of different assumed values of $V_{Z,\odot}$ on the measured warp precession rate $\omega$.

We first compute the precession rate $\omega_i$ and its MC 1$\sigma$ uncertainty $\sigma_{w,i}$ for each star, adopting a reference value of $V_{Z,\odot} = 8.6$ km s$^{-1}$. A single outlier-rejection pass is then applied, retaining only stars with $\sigma_{w,i} < 10$ km s$^{-1}$ kpc$^{-1}$. Using this cleaned sample, we scan over a range of solar vertical velocities from 6 to 12 km s$^{-1}$ in increments of 0.1 km s$^{-1}$, recomputing the precession rate for each star under each trial value. For each trial, the final precession rate is given by the statistical average $\langle\omega\rangle$ of stars beyond 12.5 kpc, and the uncertainty is expressed as the SEM: SEM$(\omega) = \frac{\sigma_{\omega}}{\sqrt{N}}$, where $\sigma_\omega^2 = \frac{1}{N-1}\sum_{i=1}^{N}(\omega_i - \langle\omega\rangle)^2$. 

\begin{figure*}
\centering
  \begin{minipage}{\textwidth}
  \includegraphics[width=\linewidth]{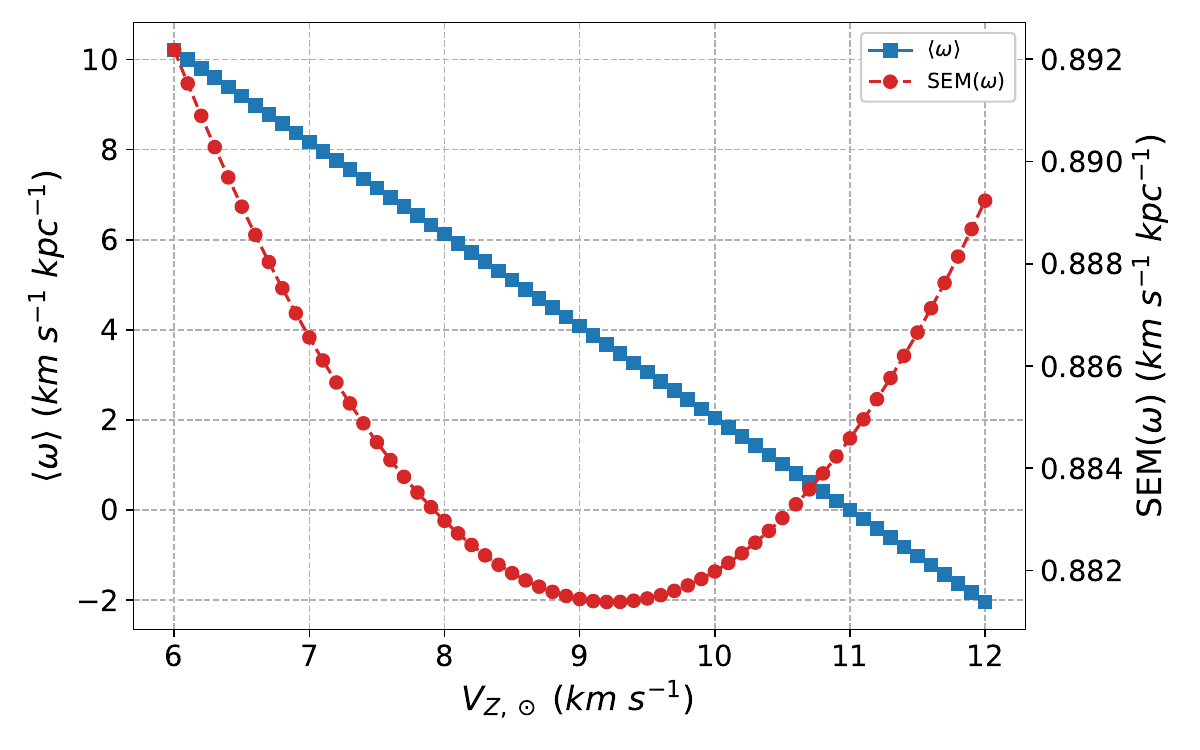}
\caption{Variation of the warp precession rate and its statistical standard error as a function of the assumed solar vertical velocity, $V_{Z,\odot}$.
Blue squares show the mean precession rate, $\langle\omega\rangle$, obtained with our updated method by averaging Cepheids at $R>12.5$ kpc; red circles give the corresponding standard error of the mean (SEM). The left-hand ordinate refers to $\langle\omega\rangle$, while the right-hand ordinate refers to the SEM.} \label{6.fig}
  \end{minipage}
  \end{figure*}

Figure~\ref{6.fig} presents the variation in results across a range of $V_{Z,\odot}$ values from 6 to 12 km s$^{-1}$. Blue squares indicate the mean precession rates, while red circles show the corresponding SEM values. The figure demonstrates an approximately linear decline in $\omega$ with increasing $V_{Z,\odot}$. Adopting the value used by \cite{2023MNRAS.523.1556D}, $V_{Z,\odot} = 6.9$ km s$^{-1}$, results in a fast precession rate of $\omega \approx 8$ km s$^{-1}$ kpc$^{-1}$, whereas our adopted value of $V_{Z,\odot} = 8.6$ km s$^{-1}$ yields a slower rate of $\omega \approx 5$ km s$^{-1}$ kpc$^{-1}$. The SEM curve shows a parabolic trend, with its minimum of 0.88 km s$^{-1}$ kpc$^{-1}$ occurring at $V_{Z,\odot} = 9.2$ km s$^{-1}$, where the corresponding $\langle\omega\rangle$ is 3.67 km s$^{-1}$ kpc$^{-1}$. This optimal value is close to our adopted one and consistent with the $V_{Z,\odot} = 9.4$ km s$^{-1}$ reported by \cite{2024arXiv241220344P}, supporting the suitability of our choice for the CC sample.

While \cite{2024arXiv241220344P} reported a nearly zero precession rate, our results do not support such a low value. As shown in Figure~\ref{6.fig}, $\omega$ approaches zero only when $V_{Z,\odot}$ increases to about 11 km s$^{-1}$. We suggest that the uncertainties in the warp precession rate derived from open clusters by \cite{2024arXiv241220344P} require more careful evaluation, particularly in accounting for the large variations in precession rates at different Galactocentric radii.

\subsection{Impact of Extinction on Warp}
\label{sec:6-2}

\cite{2025ApJ...981..179W} applied an infrared multi-band period–luminosity relations method to obtain more accurate distance estimates for Galactic CCs. The differences between their distances and those derived from Gaia samples are most prominent in the inner Galactic disk, particularly in high-extinction regions. In such regions, the use of Wesenheit magnitudes based on $G$ and $G_{\rm BP} - G_{\rm RP}$ tends to overestimate extinction and consequently underestimate distances due to extinction coefficient attenuation.

When we re-analyzed our Galactic warp model using the distances from \cite{2025ApJ...981..179W} (see Appendix~\ref{Appendix3}), we found that both the preferred model and the derived warp precession rate remained unchanged: $\omega \approx 5$ km s$^{-1}$ kpc$^{-1}$. The discrepancies between the two distance samples are mainly located within the solar circle, where the \cite{2025ApJ...981..179W} distances reveal a significantly weaker warp signature in the inner disk (see Fig.~\ref{C1.fig}). This difference likely arises because some CCs that appear to trace the inner-disk warp are actually more distant, residing at larger Galactocentric radii. This effect is particularly relevant in the directions of Galactic longitude $l \sim 75^\circ$ and $l \sim 255^\circ$, where the most suitable CCs for tracing the warp structure are concentrated.

\begin{figure*}
\centering
  \begin{minipage}{\textwidth}
  \includegraphics[width=\linewidth]{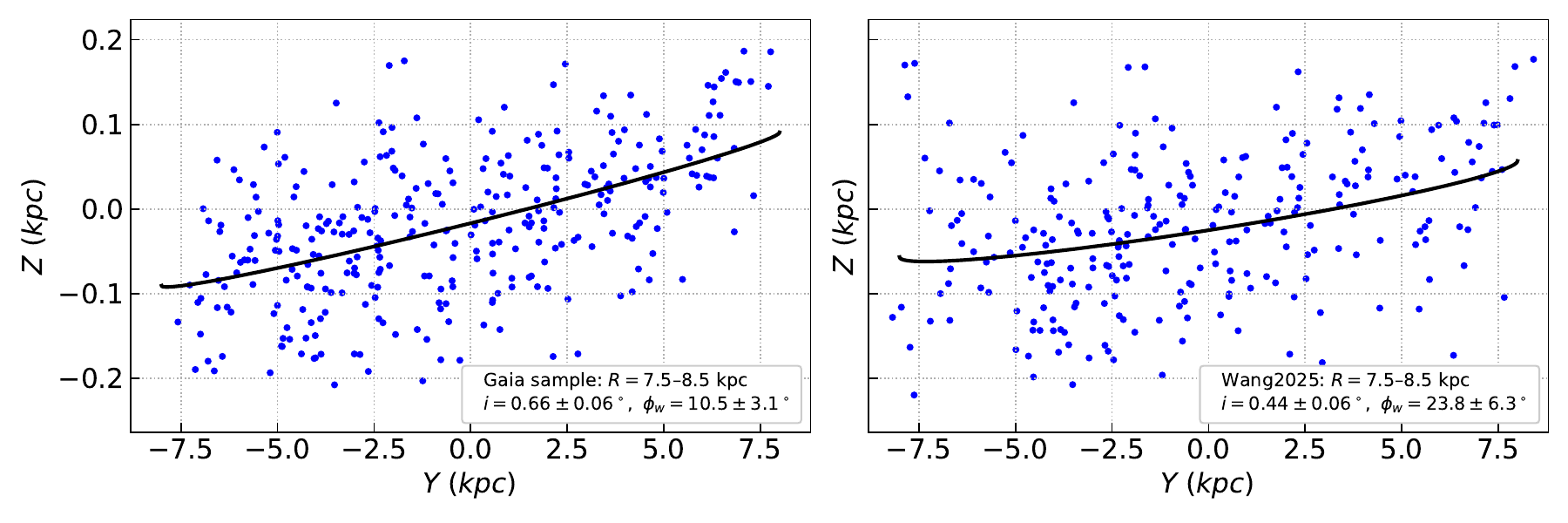}
\caption{Local solar-tilt angles obtained from different samples. The left panel shows the Gaia sample and the right panel shows the \cite{2025ApJ...981..179W} sample. Blue filled circles mark individual Cepheids; the solid black curve in each panel shows the best-fitting warp ring model $Z_{w}=R\tan(i)\sin(\phi-\phi_{w})$. The fitted parameters are listed in the in-panel legends. } \label{7.fig}
  \end{minipage}
  \end{figure*}

In the previous section, we addressed the issue of the Sun's local orbital tilt. Here, we estimate the warp inclination at $R = 8$ kpc using both samples. Figure~\ref{7.fig} compares the $Y$–$Z$ projections of CCs within the $R = 7.5$–$8.5$ kpc annulus: the left panel corresponds to the Gaia DR3 sample, and the right to the sample from \cite{2025ApJ...981..179W}. In both panels, blue points represent the stellar positions, while the black curves show the best-fitting warp ring model described by $Z_w = R\tan(i)\sin(\phi - \phi_w)$ \citep{2023MNRAS.523.1556D}, where both the local warp inclination $i$ and the LON angle $\phi_w$ are free parameters. At $R = 8$ kpc, the Gaia sample yields an inclination of $i = 0.66 \pm 0.06^\circ$, whereas the result from the \cite{2025ApJ...981..179W} sample is slightly lower at $i = 0.44 \pm 0.06^\circ$. The reduced warp amplitude in the latter is likely due to the suppression of spurious vertical displacements caused by high-extinction regions.

These results highlight the critical importance of precise extinction correction in highly obscured regions of the inner disk for accurately mapping Galactic structure. Future analyses of the warp onset radius will require more refined extinction corrections or reliance on observations in infrared bands that are less affected by dust. In contrast, the outer disk generally exhibits low extinction ($A_V < 1$ mag), leading to minimal distance differences between the two samples. Consequently, the derived structural models and warp precession rates in the outer disk are highly consistent.

\subsection{Systematic Uncertainties in the Warp Precession Rate}
\label{sec:6-3}
To quantify the systematic uncertainty in the warp precession rate, we performed a series of controlled tests, each isolating the effect of a single perturbation. In each test, one factor was varied while all other procedures were held fixed, and the mean warp precession rate was recalculated. The resulting contributions are summarized in Table \ref{tab:sys_err}, ordered by magnitude.

\begin{enumerate}
\item Solar vertical velocity $V_{Z,\odot}$. We adopted a Gaussian prior error with $1\sigma$ = 1 km s$^{-1}$, based on the range reported in the literature \citep{2024arXiv241220344P}. By perturbing $V_{Z,\odot}$ to its $\pm 3\sigma$ extremes, we recalculated the precession rate at each extreme and derived the corresponding $1\sigma$ uncertainty from their spread.
\item Warp model dependence (six models in Table \ref{tab:my_label}). For each prescription, we derived the kinematic expression by differentiation and recomputed $\omega$ using the same dataset. The standard deviation across the six values, 0.47 km s$^{-1}$ kpc$^{-1}$, is adopted as the model-choice systematic error.
\item Geometric model parameter uncertainties. Following \cite{2021ApJ...912..130C}, we performed MC sampling of the full covariance matrix for the \emph{power-law with twisted model}. We used $1\sigma$ standard deviation of the $\omega$ distribution as the systematic uncertainty for this source. 
\item Extinction treatment. We repeated the analysis using the sample from \citet{2025ApJ...981..179W}, which incorporates improved extinction corrections. We conservatively assigned half the difference between the two resulting precession rates (0.29 km s$^{-1}$ kpc$^{-1}$) as the systematic uncertainty due to the extinction treatment.

\item Methodological differences (least-squares vs. derivative method). We compared the least-squares fit result from the classical kinematic method (averaged over $R>12.5$ kpc) with the individual star calculation result from the direct derivative method. Half of the absolute difference between the two values (0.21 km s$^{-1}$ kpc$^{-1}$) was adopted as the systematic uncertainty.
\end{enumerate}
Additional potential sources of uncertainty such as using alternative assumptions for the evolution of the LONs $\phi_{0,w}(R,t)$, were found to be negligible and considered negligible are not included in the final uncertainty budget. 

\begin{table}[h]
\centering
\caption{Systematic Error Budget for the Warp Precession Rate\label{tab:sys_err}}
\begin{small}
\begin{tabular}{lccc}
\hline\hline
Error source &
$\omega_{\max}$ &
$\omega_{\min}$ &
$1\sigma$ (km s$^{-1}$ kpc$^{-1}$) \\[2pt]  
\hline
Solar vertical velocity $V_{Z,\odot}$ &  10.843 & $-1.287$ & 2.02 \\
Warp model dependence  &   6.03  &   4.83   & 0.47 \\
Geometric model parameter uncertainties     &   5.49  & $-1.97$  & 0.40 \\
Extinction treatment  &   &   & 0.29 \\
Methodological differences  &  &  & 0.21 \\
\hline
Total uncertainty                     &   &    & \textbf{2.14} \\
\hline\hline
\end{tabular}
\end{small}
\end{table}

\section{Conclusions} \label{sec:concl}

In this study, we investigated the structure of the Galactic CC warp by testing six different models, addressing key uncertainties in its geometry. By evaluating model performance across different radial ranges, we identified significant warp features in the inner disk ($5-9$ kpc), in contrast to previous works that typically assumed a flat disk within $R \lesssim 8-9$ kpc \citep{2001ApJ...556..181D, 2001ASPC..232..229D, 2019NatAs...3..320C, 2024arXiv240718659P}. While numerous studies have reported north-south asymmetry in the outer warp \citep{2019Sci...365..478S, 2020A&A...637A..96C, 2024MNRAS.528.4409C}, our analysis suggests that incorporating a second Fourier term may not provide an optimal improvement to the fit in the outer disk ($R>15$ kpc). We propose an optimal geometric warp model, in which the Galactic disk height increases continuously with radius following a power-law function, while the LONs undergo a prograde twist. This model reconciles discrepancies between previous warp models and observational constraints, particularly in the inner and outer disks, confirming that the LONs undergo significant twisting beyond 10 kpc. The optimal model is expressed as $Z_{w} = (0.00019\pm0.00003)R^{3.08\pm0.07}\sin(\phi - ((3.87\pm0.27)R - (41.79\pm3.95)))$. Compared with the warp model recently proposed by \citet{2024arXiv240718659P}, our model yields globally consistent results, yet it is formulated as a continuous, fully differentiable function with fewer parameters, making it especially well-suited for subsequent kinematic modeling.

For the warp precession rate, instead of relying on classical kinematic methods to estimate warp precession \citep{2020NatAs...4..590P, 2020ApJ...905...49C}, we directly differentiated the geometric model to derive a velocity expression for determining the precession rate from individual CCs. The results intuitively indicate a nearly uniform and low warp precession rate of $\omega = 4.86 \pm (0.88)_{stat} \pm (2.14)_{sys}$ km s$^{-1}$ kpc$^{-1}$ beyond 12.5 kpc, consistent with previous studies \citep{2021ApJ...912..130C, 2024ApJ...965..132Z, 2024arXiv240718659P}. Based on these findings, we propose a simple yet comprehensive time-dependent warp model: $Z_{w}(t) = 0.00019R^{3.08}\sin(\phi - (3.87R-41.79 + 4.86t))$. This model will provide key constraints on theoretical models of the Milky Way’s outer disk.

We further investigated the impact of adopting different values of the solar vertical velocity $V_{Z,\odot}$ and extinction corrections on the warp analysis. We found that the inferred warp precession rate $\omega$ systematically decreases with increasing $V_{Z,\odot}$. The statistical minimum in the uncertainty of $\omega$ occurs at $V_{Z,\odot} = 9.2$ km s$^{-1}$, where the precession rate reaches $\omega = 3.67$ km s$^{-1}$ kpc$^{-1}$. This result supports our adopted value of $V_{Z,\odot} = 8.6$ km s$^{-1}$ and indicates the presence of a nonzero, but moderate, warp precession. In addition, when using a sample with improved extinction corrections \citep{2025ApJ...981..179W}, we obtained a significantly reduced warp amplitude in the inner disk. This is likely due to the use of Gaia-band Wesenheit magnitudes overestimating extinction in high-extinction regions, which in turn leads to underestimated distances and causes outer-disk stars to be misidentified as inner-disk ones. In contrast, the outer disk structure and precession rate remained largely unchanged. These findings highlight the importance of accurate extinction treatment and vertical solar motion calibration in Galactic warp studies. In the future, with a more complete sample of CCs—especially those located on the far side of the disk and in high-extinction regions—the Galactic warp model can be further refined and updated.

\section*{Acknowledgements}
We thank the anonymous referee for the valuable comments. This work was supported by the National Natural Science Foundation of China (NSFC) through grants 12322306, 12173047, 12373028, 12233009 and 12133002. X. Chen and S. Wang acknowledge support from the Youth Innovation Promotion Association of the Chinese Academy of Sciences (no. 2022055 and 2023065). We also thank the support from the National Key Research and development Program of China, grants 2022YFF0503404. This work presents results from the European Space Agency (ESA) space mission Gaia. Gaia data are being processed by the Gaia Data Processing and Analysis Consortium (DPAC). Funding for the DPAC is provided by national institutions, in particular the institutions participating in the Gaia MultiLateral Agreement (MLA). The Gaia mission website is \url{https://www.cosmos.esa.int/gaia}. The Gaia archive website is \url{https://archives.esac.esa.int/gaia}.

\renewcommand{\bibname}{References}
\bibliographystyle{aasjournal}
\bibliography{ref}

\appendix  

\section{Excluding outliers}
\label{Appendix}

Figures \ref{A1.fig}–\ref{A3.fig} present the residual distributions obtained by fitting the initial warp model across the radial range of  $5-20$ kpc. These residuals provide insight into the deviations between the observed CC heights and the initial model, helping to exclude outliers and identify potential asymmetries. In Figures \ref{A1.fig}–\ref{A3.fig}, the y-axis ranges in the right panels represent the residual filtering threshold for each radius, with the red dashed line indicating $\Delta Z=0$. The left panels show the retained CCs (blue dots) and the excluded outliers (red dots) at each radius. The figure also displays the distribution of CCs' heights as a function of azimuth, with the solid red line representing the initial model. A significant asymmetry in residuals is observed even in the $5-9$ kpc range.

\setcounter{figure}{0}
\renewcommand*\thefigure{A\arabic{figure}}

\begin{figure*}
\centering
  \begin{minipage}{\textwidth}
  \includegraphics[width=\linewidth]{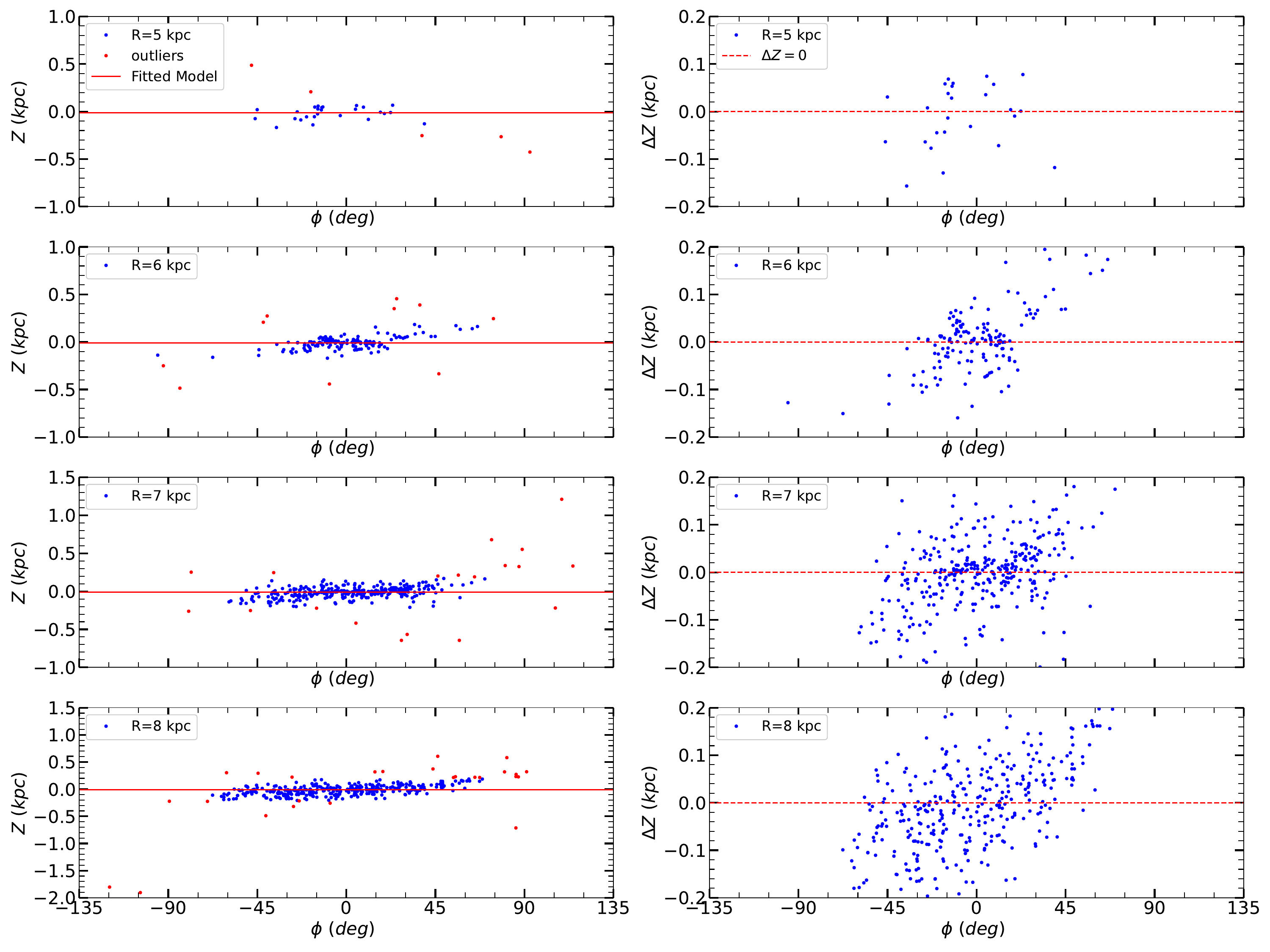}
  \caption{Outliers were excluded from the residuals for the $5-8$ kpc samples. The distributions of heights and residuals as a function of azimuth are shown separately for each radius. Blue solid dots indicate retained samples, while red solid dots indicate identified outliers. The solid red line indicates the fitted model, and red dashed line marks $\Delta Z = 0$. The range of residuals shows our selection criterion. \label{A1.fig}}
  \end{minipage}
  \end{figure*}
  
\begin{figure*}
\centering
  \begin{minipage}{\textwidth}
  \includegraphics[width=\linewidth]{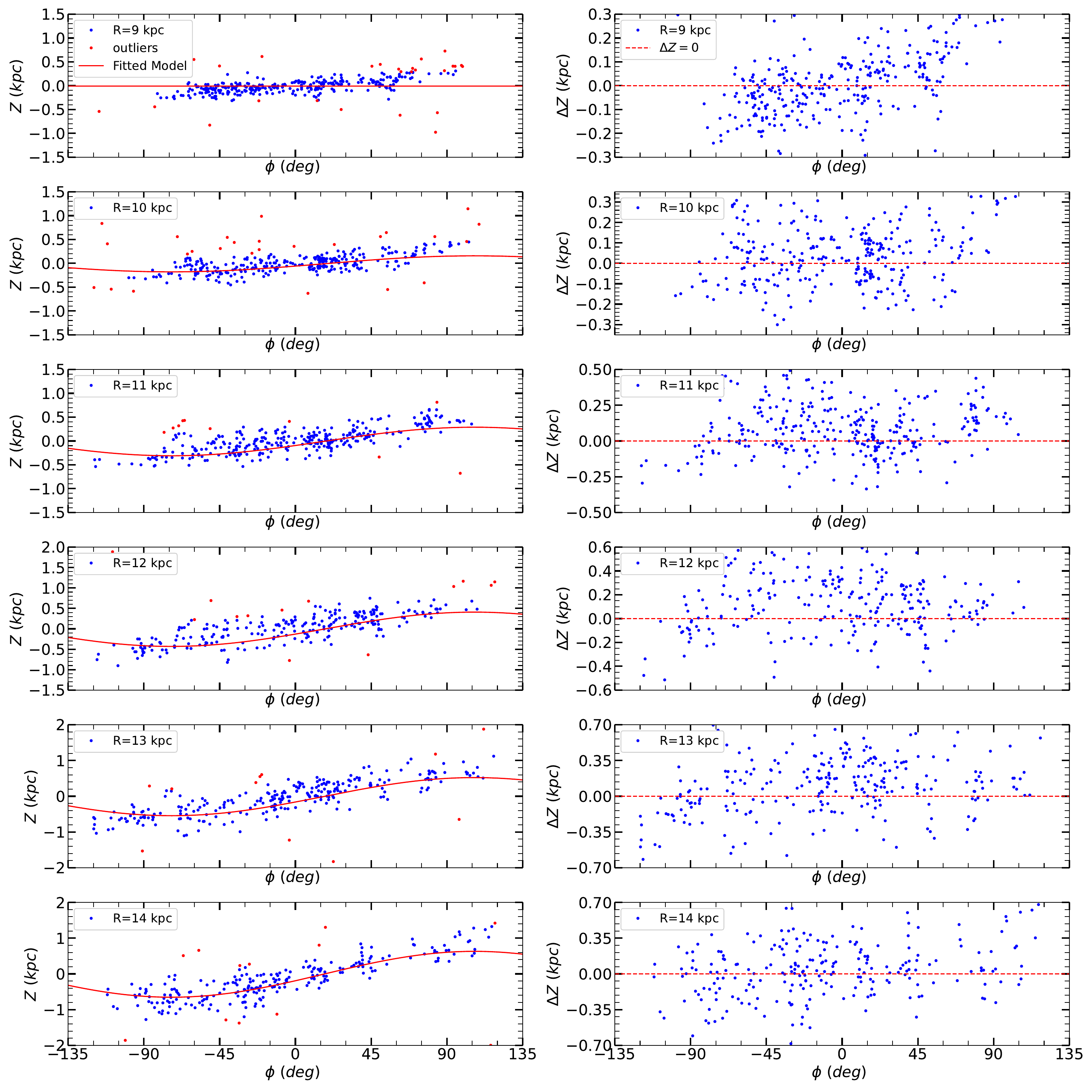}  
  \caption{Similar to Fig. \ref{A1.fig} but for $9-14$ kpc. \label{A2.fig}}
  \end{minipage}
\end{figure*}

\begin{figure*}
\centering
  \begin{minipage}{\textwidth}
  \includegraphics[width=\linewidth]{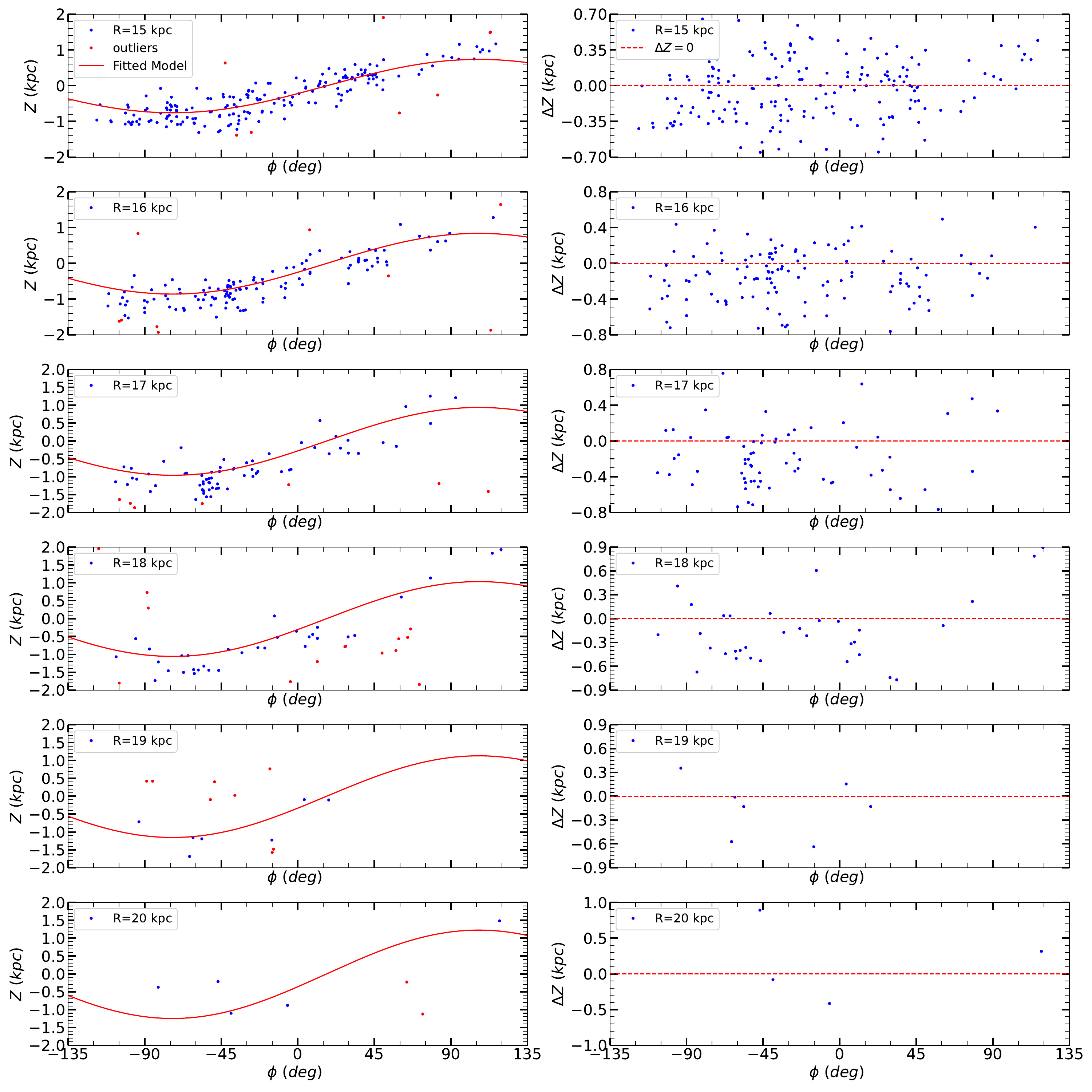}
  \caption{Similar to Fig. \ref{A1.fig} but for $15-20$ kpc.  \label{A3.fig}}
  \end{minipage}
\end{figure*}

\section{Comparison of RMSE of two twisted models.}
\label{Appendix2}

We compare the RMSE of two twisted models and the non-twisted model (\emph{Power-law model}) in Fig. \ref{B1.fig}. Both twisted models exhibit lower RMSE values beyond 10 kpc, indicating that they provide a better fit to the sample compared to the non-twisted model. The RMSE differences between the two twisted models remain minimal for $R>10$ kpc, suggesting that both effectively describe the outer disk. However, for $R<9$ kpc, the \emph{Linear and $R_{1}=9$ with twisted model} assumes a flat inner disk, leading to higher RMSE values than the \emph{Power-law with twisted model}. As discussed earlier, a warp model is more suitable for the sample within 9 kpc. Therefore, we adopt the \emph{Power-law with twisted model} as the best-fit representation of the Galactic warp.

\setcounter{figure}{0}
\renewcommand*\thefigure{B\arabic{figure}}

\begin{figure*}
\centering
  \begin{minipage}{\textwidth}
  \includegraphics[width=\linewidth]{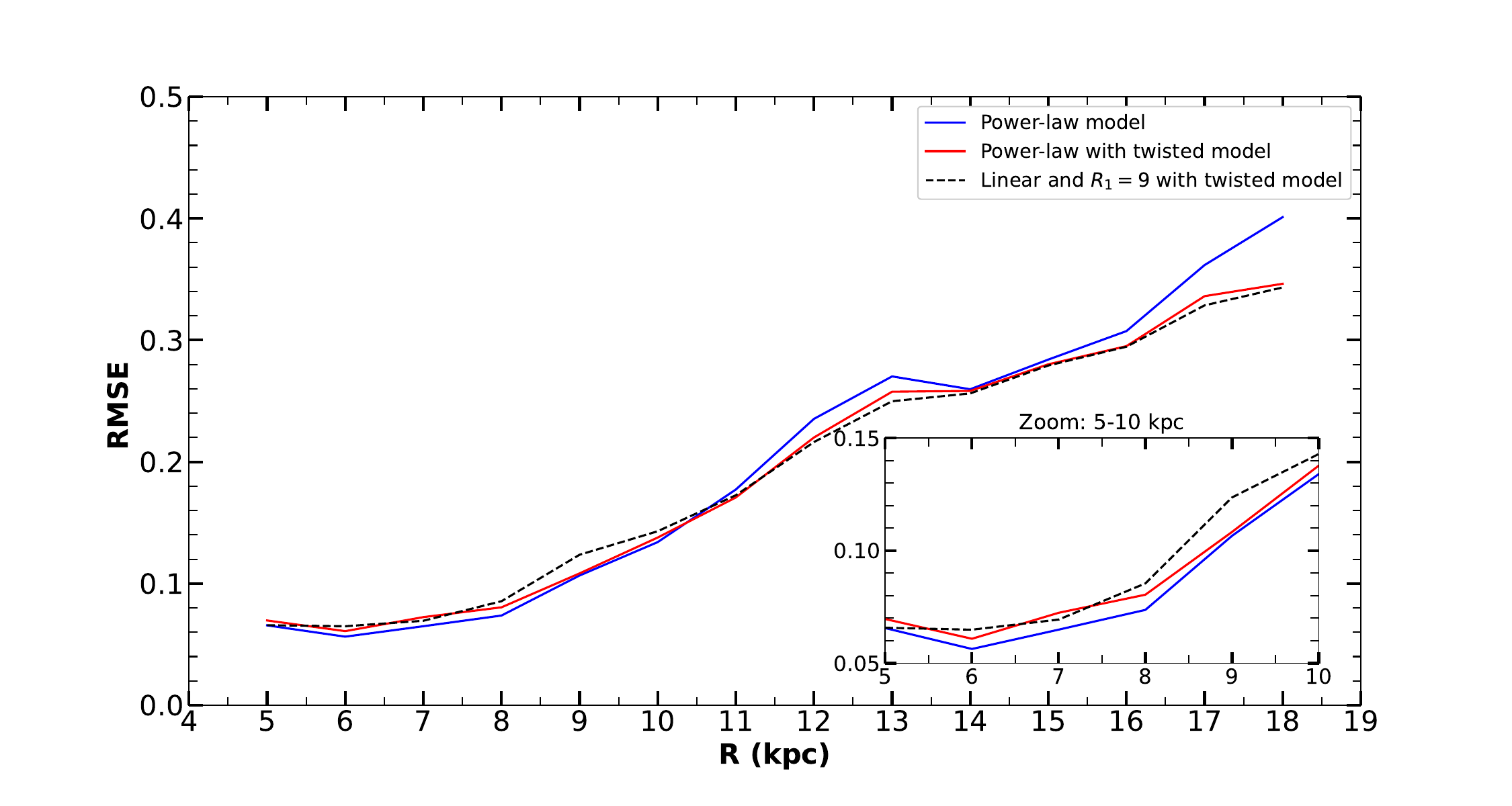}
\caption{Similar to Fig. \ref{1.fig}, but comparing the two twisted models with the power-law model. The \emph{Power-law model}, \emph{Power-law with twisted model}, and \emph{Linear and $R_{1}=9$ with twisted model} are represented by blue solid, red solid and black dotted lines, respectively. \label{B1.fig}}
  \end{minipage}
  \end{figure*}

\section{Results Based on the Infrared-Calibrated Cepheid Sample}
\label{Appendix3}

Following the methodology outlined in the main text, we re-applied our full processing pipeline to the CC catalog of \cite{2025ApJ...981..179W}, which provides heliocentric distances derived from multi-band infrared photometry. These distances are expected to be more accurate than those based on Gaia-band data, particularly in regions of high extinction. The primary difference between this catalog and the Gaia-based sample lies in the inner Galactic disk. A comparison between Fig. \ref{C1.fig} and Fig. \ref{A1.fig} reveals that within the $5-9$ kpc range, the warp signature in the \cite{2025ApJ...981..179W} sample is substantially weaker, with a notably reduced amplitude. Despite this difference, model fitting using the new sample confirms the same optimal warp formulation as identified previously, expressed as: $Z_{w} = (0.00012\pm0.00002)R^{3.25\pm0.07}\sin(\phi - ((3.84\pm0.29)R - (40.16\pm4.28)))$. Compared to the model derived from the Gaia sample, the main difference lies in the lower warp amplitude in the inner disk, but the results remain consistent within the $1\sigma$ uncertainties. Substituting this warp geometric model into the kinematic framework yields a warp precession rate of $\omega = 5.43\pm0.93$ km s$^{-1}$ kpc$^{-1}$ for stars beyond 12.5 kpc. This result is consistent with our main-text estimate from Gaia sample, demonstrating the robustness of both the global warp geometry and the warp precession rate measurements across the two datasets.

\setcounter{figure}{0}
\renewcommand*\thefigure{C\arabic{figure}}

\begin{figure*}
\centering 
  \begin{minipage}{\textwidth}
  \includegraphics[width=\linewidth]{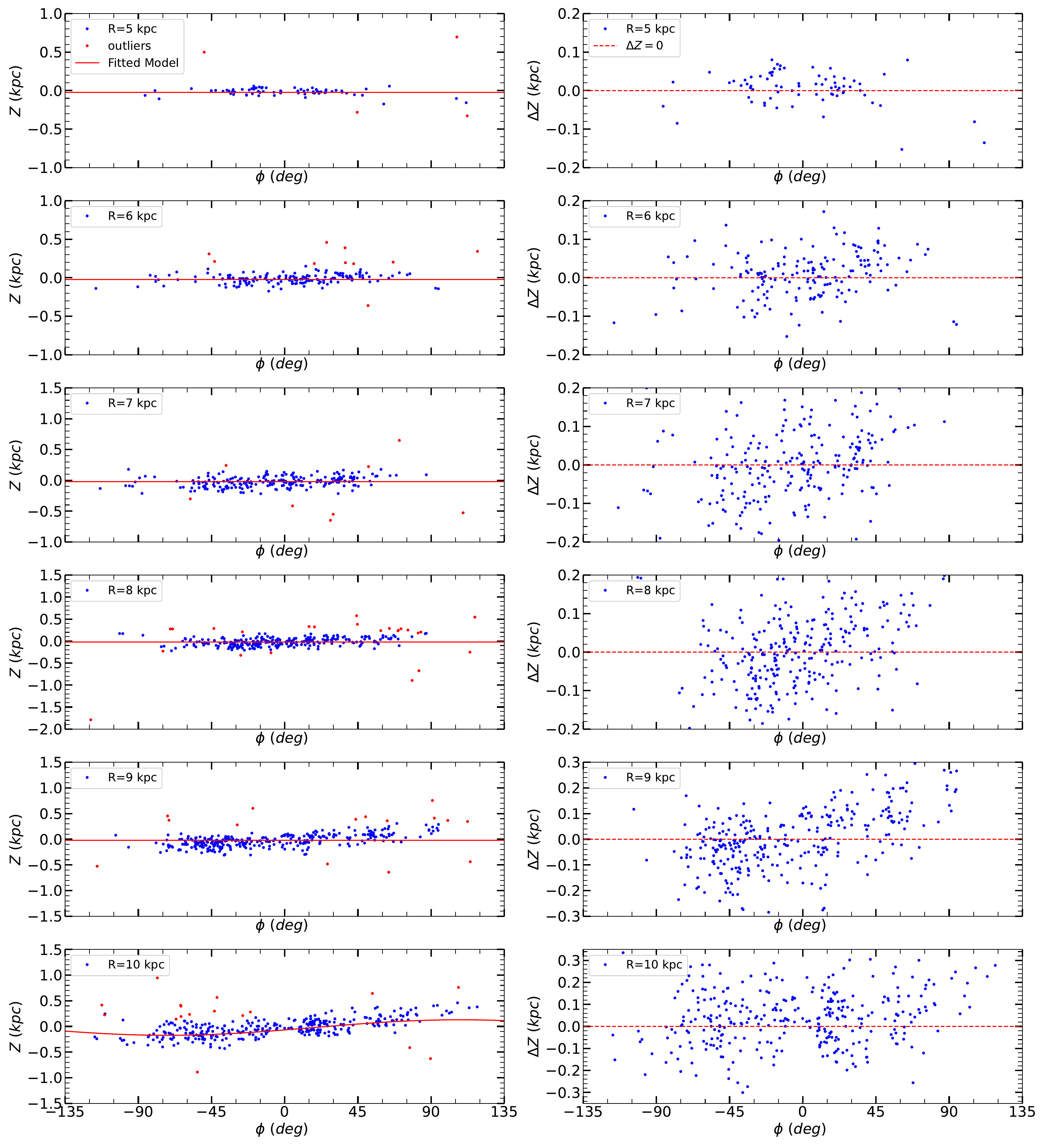}
  \caption{Similar to Appendix \ref{A1.fig}, but for \cite{2025ApJ...981..179W} samples $5-10$ kpc. \label{C1.fig}}
  \end{minipage}
  \end{figure*}

\end{document}